\documentclass[twocolumn,twoside,preprintnumbers,amsmath,amssymb,pacs]{revtex4}
\usepackage{epsfig}
\usepackage{graphicx}
\usepackage{dcolumn}
\usepackage{bm}

\usepackage{fancyhdr}

\usepackage{pslatex}

\begin{document}

\title{{\Large Nucleation in the chiral transition with an 
inhomogeneous background}}

\author{Bruno G. Taketani and Eduardo S. Fraga}

\affiliation{Instituto de F{\'\i}sica, Universidade 
Federal do Rio de Janeiro \\
C.P.~68528, Rio de Janeiro, RJ 21941-972, Brazil}

\received{on 24 March, 2006}

\begin{abstract}
We consider an approximation procedure to evaluate 
the finite-temperature one-loop fermionic density 
in the presence of a chiral background field which systematically 
incorporates effects from inhomogeneities in the chiral field through a 
derivative expansion. Modifications in the effective potential and their 
consequences for the bubble nucleation process are discussed.

PACS numbers:

Keyword:
\end{abstract}

\maketitle

\thispagestyle{fancy}
\setcounter{page}{0}



The mechanism of chiral symmetry breaking and the 
study of the dynamics of phase conversion after a temperature-driven chiral 
transition can be done conveniently within low-energy effective models 
\cite{quarks-chiral,ove,Scavenius:1999zc,Scavenius:2000qd,
Scavenius:2001bb,paech,Aguiar:2003pp,polyakov,explosive}. 
In particular, to study the mechanisms of bubble nucleation and 
spinodal decomposition in a hot expanding plasma \cite{Csernai:1992tj}, 
it is common to adopt the linear $\sigma$-model
coupled to quarks \cite{gellmann}. 
The gas of quarks provides a thermal bath in which the long-wavelength
modes of the chiral field evolve, and the latter plays the role of an order
parameter in a Landau-Ginzburg approach to the description of the chiral
phase transition. 
The standard procedure is then integrating over the fermionic degrees of 
freedom, using a classical approximation for the chiral field, to obtain a formal 
expression for the thermodynamic potential from which one 
can obtain all the physical quantities of interest. 

To compute correlation functions and thermodynamic 
quantities, one has to evaluate the fermionic determinant that results 
from the functional integration over the quark fields within some 
approximation scheme. Usually one performs a standard one-loop 
calculation assuming a homogeneous and static background 
field \cite{kapusta-book}. Nevertheless, for a system that is in the process 
of phase conversion after a chiral transition, one expects inhomogeneities 
in the chiral field configuration due to fluctuations to play a major 
role in driving the system to the true ground state. In fact, for high-energy 
heavy ion collisions, hydrodynamical studies have shown that significant 
density inhomogeneities may develop dynamically when the chiral transition 
to the broken symmetry phase takes place \cite{paech}, and inhomogeneities 
generated during the late stages of the nonequilibrium evolution of the 
order parameter might leave imprints on the final spatial distributions 
and even on the integrated, inclusive abundances \cite{licinio}.

In this paper we briefly describe an approximation procedure to evaluate 
the finite-temperature fermionic density in the presence 
of a chiral background field, which systematically incorporates 
effects from inhomogeneities in the bosonic field through a 
gradient expansion. (For details, see Ref. \cite{BT-ESF}.) 
The method is valid for the case in which 
the chiral field varies smoothly, and allows one to extract information 
from its long-wavelength behavior, incorporating corrections order 
by order in the derivatives of the field. For simplicity, we ignore 
corrections coming from bosonic fluctuations in what follows. Those 
would result in a bosonic determinant correction to the effective 
potential \cite{bosonic}. To focus on the effect of an inhomogeneous 
background field on the fermionic density, we treated the scalar 
field essentially as a  ``heavy''
(classical) field, whereas fermions were assumed to be ``light''.



Let us consider a scalar field $\phi$ coupled to fermions $\psi$ 
according to the Lagrangian 
\begin{equation}
\mathcal{L} = \overline{\psi}[i\gamma^{\mu}\partial _{\mu} + \mu\gamma^0 -
M(\phi)]\psi + \frac{1}{2}\partial_{\mu}\phi ~\partial^{\mu}\phi - V(\phi)\; ,
\label{lagrangian}
\end{equation}
where $\mu$ is the fermionic chemical potential, $M(\phi)$ is the effective 
mass of the fermions and $V(\phi)$ is a self-interaction potential for the 
bosonic field.

In the case of the linear $\sigma$-model coupled to quarks, $\phi$ represents 
the $\sigma$ direction of the chiral field $\Phi=(\sigma,\vec{\pi})$, where 
$\pi^i$ are pseudoscalar fields playing the role of the pions, which we drop 
here for simplicity. The field $\psi$ plays the role of the constituent-quark field 
$q=(u,d)$, and $\mu=\mu_q$ is the quark chemical potential. The 
``effective mass'' is given by $M(\phi)=g|\sigma|$, 
and $V(\Phi)=(\lambda^2/4)(\sigma^2+\vec{\pi}^2-v^2)^2-h_q\sigma$ is the 
self-interaction potential for $\Phi$. The parameters above are
chosen such that chiral $SU_{L}(2) \otimes SU_{R}(2)$ symmetry is
spontaneously broken in the vacuum. The vacuum expectation values of the
condensates are 
$\langle\sigma\rangle =\mathit{f}_{\pi}$ and $\langle\vec{\pi}\rangle =0$, 
where $\mathit{f}_{\pi}=93$~MeV is the pion decay constant.
The explicit symmetry breaking term is due to the finite current-quark
masses and is determined by the PCAC relation, giving 
$h_q=f_{\pi}m_{\pi}^{2}$, where $m_{\pi}=138$~MeV is the pion mass. This
yields $v^{2}=f^{2}_{\pi}-{m^{2}_{\pi}}/{\lambda ^{2}}$. The value of 
$\lambda^2 = 20$ leads to a $\sigma$-mass, 
$m^2_\sigma=2\lambda^{2}f^{2}_{\pi}+m^{2}_{\pi}$, equal to 600~MeV. 

The Euler-Lagrange equation for static chiral field configurations 
contains a term which represents the fermionic density:
\begin{equation}
\nabla^2\phi=\frac{\partial V}{\partial\phi}+g\rho(T,\mu,\phi)\; ,
\label{euler-lagrange}
\end{equation}
and the density of fermions at a given point $\vec{x}_0$ has the form 
\begin{equation}
\rho(\vec{x}_0)=Sp \left\langle\vec{x}_0 
\left\vert \frac{1}{G_E^{-1}+M(\hat{x})}\right\vert 
\vec{x}_0\right\rangle \; ,
\label{density-Sp}
\end{equation}
where $\vert\vec{x}_0\rangle$ is a position eigenstate with eigenvalue 
$\vec{x}_0$, and $Sp$ represents a trace over fermionic degrees of 
freedom, such as color, spin and isospin. 



In order to take into account inhomogeneity effects of the chiral background 
field, $\phi$, encoded in the position dependence of $M$ in (\ref{density-Sp}), 
we resort to a derivative expansion as explained below \cite{BT-ESF}.

In momentum representation, the expression for the 
fermionic density assumes the form
\begin{equation}
\rho(\vec{x}_0)=Sp~ 
T\sum_n \int \frac{d^3k}{(2\pi)^3}
e^{-i\vec{k}\cdot\vec{x}} ~
\frac{1}{\gamma^0 (i\omega_n+\mu)-\vec{\gamma}\cdot\vec{k}+M(\hat{x})}
~e^{i\vec{k}\cdot\vec{x}} \; .
\end{equation}
One can transfer the $\vec{x}_0$ dependence to $M(\hat{x})$ through a 
unitary transformation, obtaining
\begin{equation}
\rho(\vec{x}_0)=Sp~ 
T\sum_n \int \frac{d^3k}{(2\pi)^3}
\frac{1}{\gamma^0 (i\omega_n+\mu)-\vec{\gamma}\cdot\vec{k}
+M(\hat{x}+\vec{x}_0)} \; ,
\end{equation}
where one should notice that $\vec{x}_0$ is a c-number, not an operator.

Now we expand $M(\hat{x}+\vec{x}_0)$ around $\vec{x}_0$:
\begin{eqnarray}
M(\hat{x}+\vec{x}_0)&\equiv& 
M(\vec{x}_0)+\Delta M(\hat{x},\vec{x}_0)=\nonumber \\ 
&& M(\vec{x}_0)+\nabla_i M~\hat{x}^i +
\frac{1}{2}\nabla_i \nabla_j M~\hat{x}^i \hat{x}^j + 
\cdots \; ,
\end{eqnarray}
and use $\hat{x}^i=-i\nabla_{k_i}$ to write
\begin{widetext}
\begin{equation}
\rho(\vec{x}_0)=Sp~ T\sum_n \int \frac{d^3k}{(2\pi)^3} 
\frac{1}{\gamma^0(i\omega_n+\mu)-\vec{\gamma}\cdot\vec{k}+M(\vec{x}_0)}
\left[ 1+ \Delta M(-i\nabla_{k_i},\vec{x}_0) 
\frac{1}{\gamma^0(i\omega_n+\mu)-\vec{\gamma}\cdot\vec{k}+M(\vec{x}_0)}
\right]^{-1}\, .
\label{brackets}
\end{equation}
\end{widetext}

To study the dynamics of phase conversion after a chiral 
transition, one can focus on the long-wavelength properties 
of the chiral field. From now on we assume that the 
static background, $M(\vec{x})$, varies smoothly and fermions 
transfer a small ammount of momentum to the chiral field. Under 
this assumption, we can expand the expression inside brackets 
in Eq. (\ref{brackets}) in a power series:
\begin{widetext}
\begin{equation}
\rho(\vec{x})=Sp~ T\sum_n \int \frac{d^3k}{(2\pi)^3} 
\frac{1}{\gamma^0(i\omega_n+\mu)-\vec{\gamma}\cdot\vec{k}+M(\vec{x})}
~\sum_\ell (-1)^\ell 
\left[ \Delta M(-i\nabla_{k_i},\vec{x}) 
\frac{1}{\gamma^0(i\omega_n+\mu)-\vec{\gamma}\cdot\vec{k}+M(\vec{x})}
\right]^{\ell}\, . 
\label{expansion}
\end{equation}
\end{widetext}
Eq. (\ref{expansion}), together with
\begin{equation}
\Delta M(-i\nabla_{k_i},\vec{x})=
\nabla_i M\left(\frac{1}{i}\right)\nabla_{k_i}+
\frac{1}{2}\nabla_i \nabla_j M\left(\frac{1}{i}\right)^2 
\nabla_{k_i}\nabla_{k_j} + \cdots \; ,
\end{equation}
provides a systematic procedure to incorporate corrections brought about 
by inhomogeneities in the chiral field to the quark density, so that 
one can calculate 
$\rho(\vec{x})=\rho_0(\vec{x})+\rho_1(\vec{x})+\rho_2(\vec{x})+\cdots$ 
order by order in powers of the derivative of the background, $M(\vec{x})$.




The leading-order term in this gradient expansion for $\rho(\vec{x})$ 
can be calculated in the standard fashion \cite{kapusta-book} and yields 
the well-known mean field result for the scalar quark density
\begin{equation}
\rho_0 =  \nu_q \int \frac{d^3k}{(2\pi)^3} 
\frac{M(\phi)/E_k(\phi)}{e^{[E_k(\phi)-\mu_q]/T}+1} + 
(\mu_q \to -\mu_q) \; ,  
\label{rho0}
\end{equation}
where $\nu_q=12$ is the color-spin-isospin degeneracy factor, 
$E_k(\phi)=(\vec{k}^2+M^2(\phi))^{1/2}$, and 
$M(\phi)=g|\phi|$ plays the role of an effective mass for the quarks. 
The net effect of this leading term is correcting the potential for the chiral 
field, so that we can rewrite Eq. (\ref{euler-lagrange}) as
\begin{equation}
\nabla^2\phi=\frac{\partial V_{eff}}{\partial\phi} \; ,
\label{euler-lagrange2}
\end{equation}
where $V_{eff}= V(\phi)+V_q(\phi)$ and
\begin{equation}
V_q\equiv -\nu_q T \int \frac{d^3k}{(2\pi)^3} 
\ln\left( e^{[E_k(\phi)-\mu_q]/T}+1 \right)+
(\mu_q \to -\mu_q) \; .
\end{equation}

This kind of effective potential is commonly used as the coarse-grained 
thermodynamic potential in a phenomenological 
description of the chiral transition for an expanding 
quark-gluon plasma 
\cite{Scavenius:1999zc,Scavenius:2000qd,Scavenius:2001bb,paech}. 



%
\begin{figure*}[htbp]
  \vspace{0.5cm}
  \centerline{\hbox{ \hspace{-0.2in}
    \includegraphics[angle=0,width=3in]{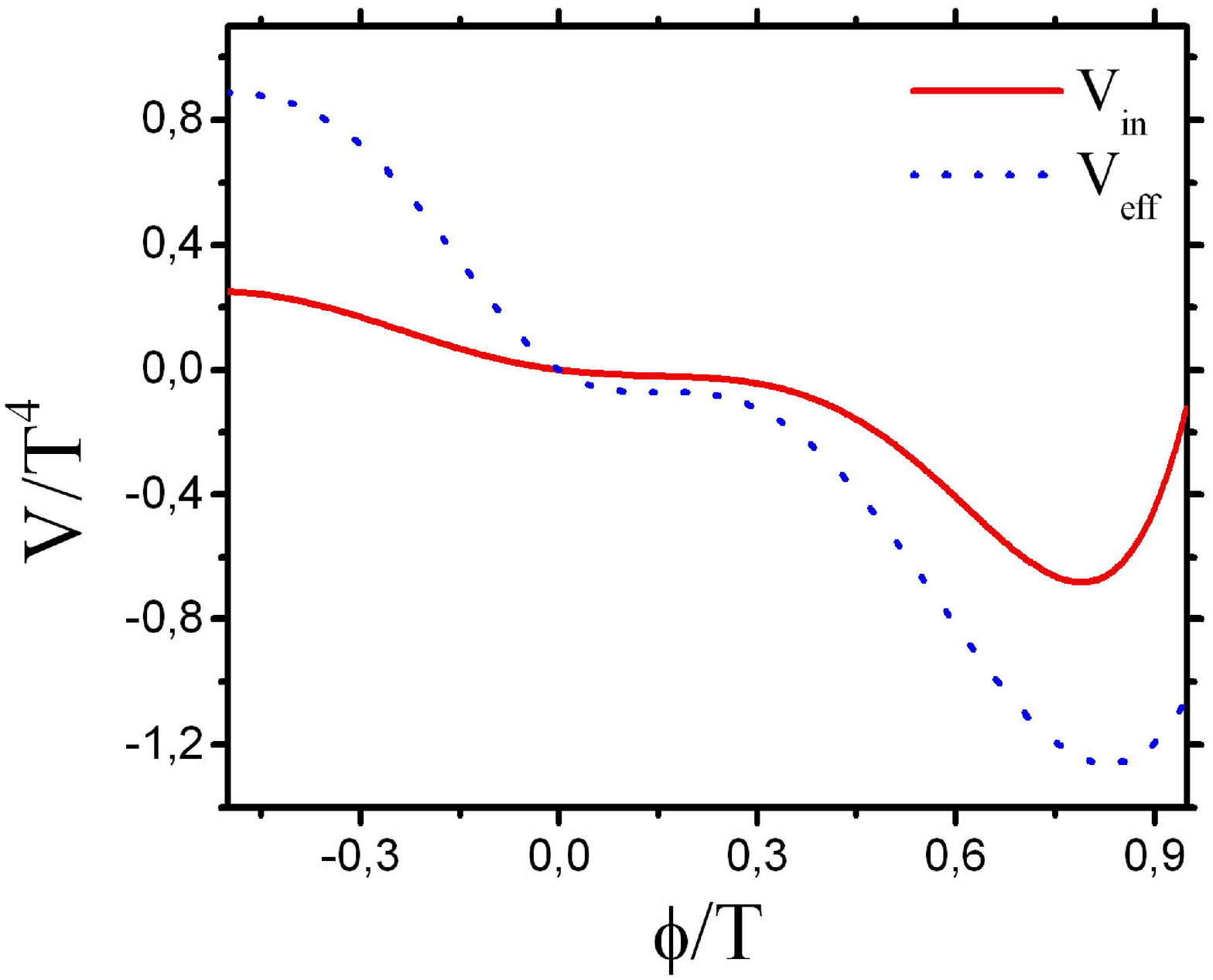}
    \hspace{0.25in}
    \includegraphics[angle=0,width=3in]{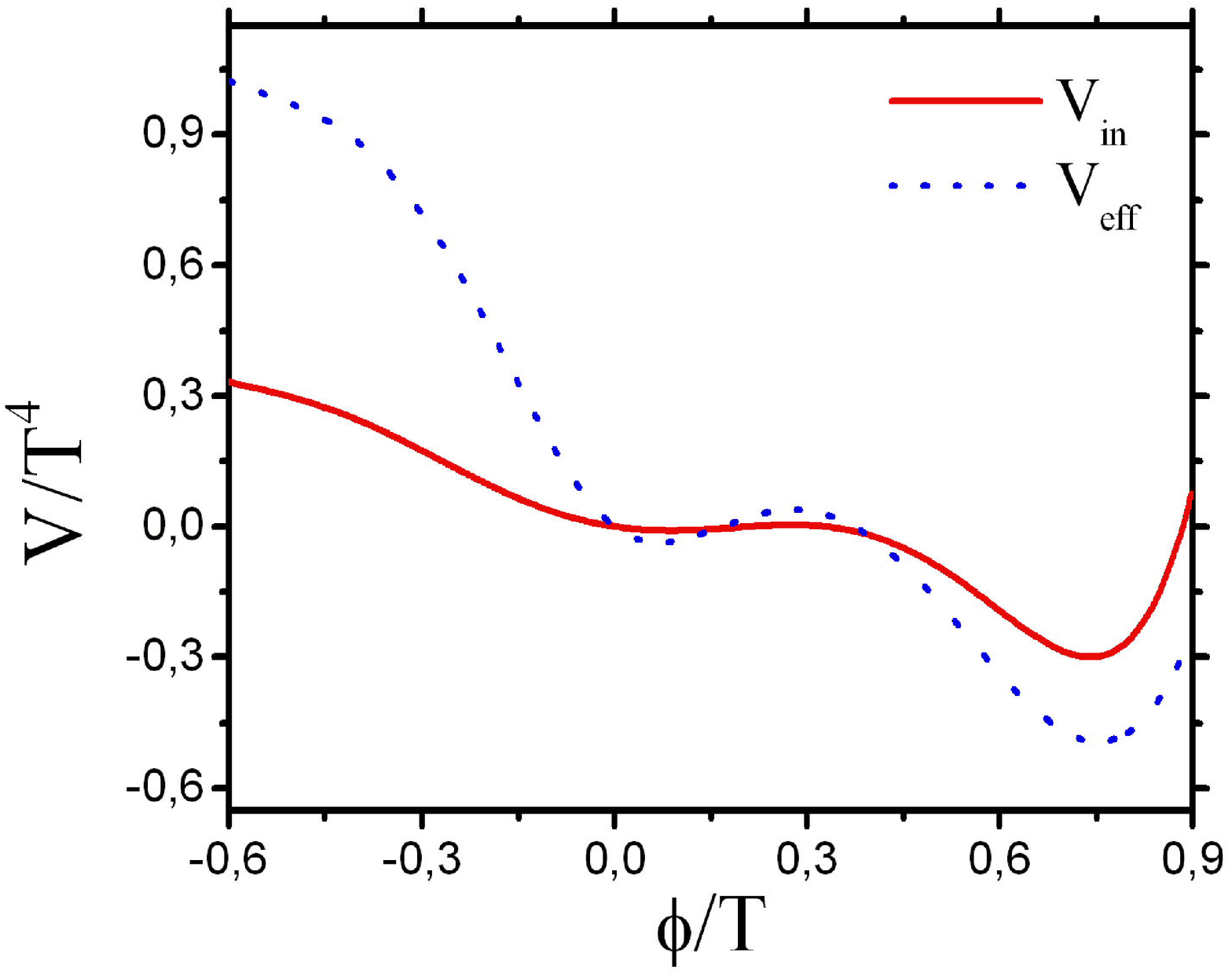}
    }
  }
  \vspace{-10pt}
  \hbox{\hspace{1.4in} (a) \hspace{2.97in} (b) }
  \vspace{9pt}

  \centerline{\hbox{ \hspace{-0.2in}
    \centerline{\includegraphics[angle=0,width=3in]{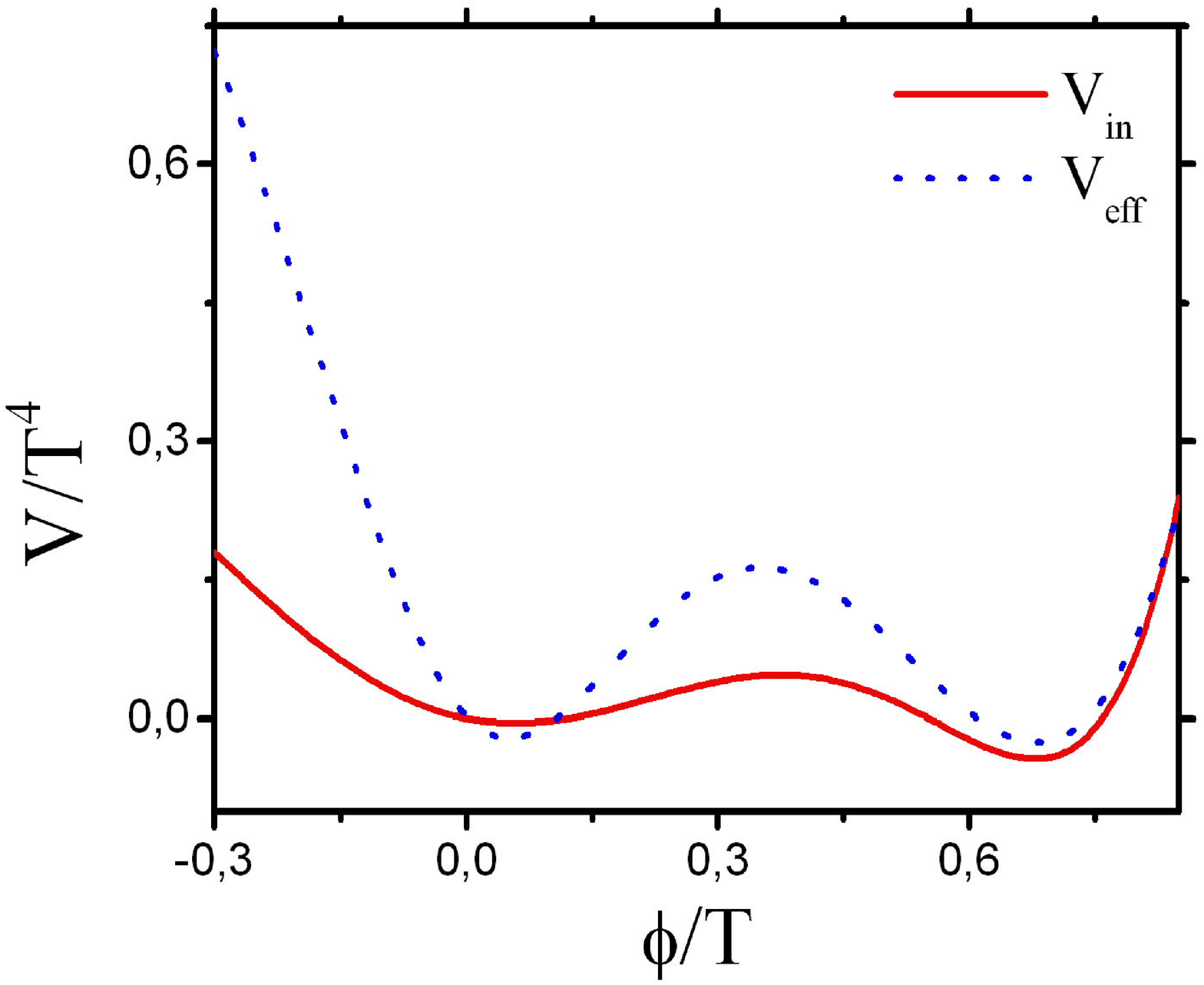}}
    \hspace{-3.8in}
    \centerline{\includegraphics[angle=0,width=3in]{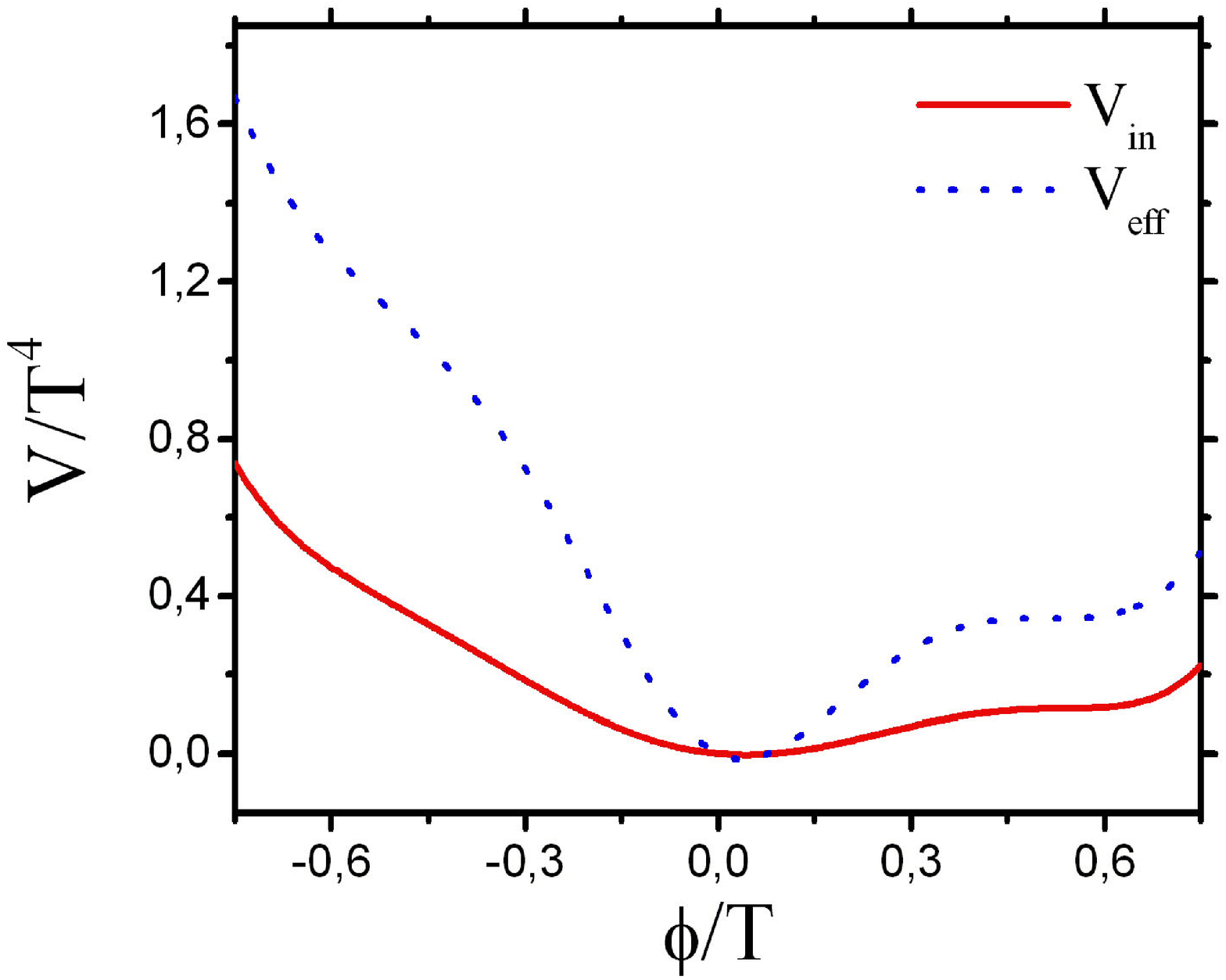}}
    }
  }
  \vspace{-160pt}
  \vspace{135pt}

  \vspace{15pt}
  \hbox{\hspace{1.4in} (c) \hspace{2.97in} (d) }
  \vspace{9pt}
  \label{g=3.3}

  \caption{$V_{eff}$ and $V_{in}(\phi)$ for different 
values of the temperature $T=[108~ {\rm (a)}, 116~ {\rm (b)}, 
124~ {\rm (c)}, 132~ {\rm (d)}]$ MeV 
at $\mu_q=0$ and for $g=5.5$.}
\label{Veff-Vin}
\end{figure*}

The next non-trivial term in the expansion contains two 
contributions: one coming from $\rho_1$ and another from $\rho_2$. 
This is due to the rearrangement of powers of the gradient 
operator. This term will correct the Laplacian piece in the chiral 
field equation. Dropping zero-temperature contributions which can 
be absorbed by a redefinition of the bare parameters in $V$, 
a long but straightforward calculation yields \cite{BT-ESF}
\begin{equation}
(\rho_1 + \rho_2)=-(\nabla^2 M)~W_q(T,\mu_q,\phi)\; ,
\label{rho1+rho2}
\end{equation}
where 
\begin{equation}
W_q(T,\mu_q,\phi)=\frac{\nu_q}{2\pi^2}\int_0^\infty~
dk k^2 [H(E_k,T,\mu_q) 
+ H(E_k,T,-\mu_q)] \; ,
\end{equation}
and $H(E_k,T,\mu_q)$ is a complicated function whose particular form 
is not very illuminating. In the low-temperature limit ($\beta M >> 1$), 
the integral above is strongly suppressed 
for high values of $k$, and the leading term has the much simpler form
\begin{equation}
W_q(T,\mu_q,\phi)\approx \frac{\nu_q}{\pi^2}\frac{\sqrt{2\pi}}{8}
e^{-\beta|M|} (\beta |M|)^{3/2} \, ,
\end{equation}
which gives a a better idea of the profile of the first inhomogeneity 
correction. One can already anticipate that it will be concentrated 
in the same region where the homogeneous correction was 
significant, i.e. $\beta M < 1$ (cf. \cite{Scavenius:2001bb}), 
being exponentially suppressed for higher values of the field. In 
fact, a numerical study of the complete $W_q$ shows that this 
function is peaked around $\phi=0$ and non-negligible for 
$\beta|\phi|<1$ \cite{BT-ESF}.

The Euler-Lagrange equation for the chiral field up to this order in the gradient 
expansion reads
\begin{equation}
\nabla^2\phi=[1+gW_q(T,\mu_q,\phi)]^{-1}~
\frac{\partial V_{eff}(T,\mu_q,\phi)}{\partial\phi} \equiv 
V'_{in}(\phi)\; .
\end{equation}

The complete new effective potential can be obtained from our 
previous results by numerical integration. In order to proceed 
analytically, though, we choose to fit its derivative, which we 
know exactly up to this order, by a polynomial of the fifth degree. 
Actually, we know that the commonly used effective potential, 
$V_{eff}=V+V_q$, can hardly be distinguished from a fit with a 
polynomial of sixth degree in the region of interest for nucleation 
\cite{Scavenius:2001bb,Fraga:2004hp}. Working with fits will 
be most convenient for using well-known results in the thin-wall 
approximation to estimate physical quantities that are relevant 
for nucleation, such as the surface tension and the free energy 
of the critical bubble \cite{reviews}. 

We can now integrate analytically the polynomial approximation 
to the derivative of the complete effective potential. In Fig. 1 we 
display the curves for $V_{eff}$ and $V_{in}$ for a few values 
of temperature and $\mu_q=0$. From the figure one can notice a few 
consequences of the inhomogeneity correction. A first general 
effect is the smoothening of the effective 
potential. In particular, and most importantly, the barrier between 
the symmetric phase and the broken phase is significantly diminished, 
as well as the depth of the broken phase minimum. Therefore, one can 
expect an augmentation in the bubble nucleation rate. In principle, one 
should also have better results from calculations within the thin-wall 
approximation. Also, the critical temperature moves up slightly. 



Let us now consider the effects of the first inhomogeneity correction on 
the process of phase conversion driven by the nucleation of bubbles 
\cite{reviews}. To work with approximate analytic formulas, we follow 
Ref. \cite{Scavenius:2001bb} and express $V_{in}$ over the range 
$0 \le \phi \le T$ in the familiar Landau-Ginzburg form
$V_{\rm eff} \approx \sum_{n=0}^4 a_n \, \Phi^n$.
Although this approximation is obviously incapable of reproducing all three 
minima of $V_{in}$, this polynomial form is found to provide a good quantitative 
description of $V_{in}$ in the region of interest for nucleation, i.e. where 
the minima for the symmetric and broken phases, as well as  the barrier 
between them, are located. Using the thin-wall approximation, we can 
compute the relevant quantities in the way described in Ref. 
\cite{Scavenius:2001bb}. Nevertheless, it was shown in \cite{Scavenius:2001bb} 
that the thin-wall limit becomes very imprecise as one approaches the 
spinodal. In this vein, the analysis presented below is to be regarded as 
semi-quantitative. To be consistent will compare results from the homogeneous 
calculation to those including the inhomogeneity correction within the 
same approximation \cite{BT-ESF}. 

To illustrate the effect, we compute 
the critical radius, $R_c$, and free energy of the critical bubble in units of 
temperature, $F_b/T$, for three different values of the temperature. 
For $T=108~$MeV, corresponding to the spinodal, which is 
not modified by the first inhomogeneity correction, the corrected 
values are $R_c\approx 0.98~$fm and $F_b/T\approx 0.20$, as compared to 
$R_c\approx 1.1~$fm and $F_b/T\approx 0.9$ in the homogeneous case. 
The same computation for $T=116~$MeV yields 
$R_c\approx 2.15~$fm and $F_b/T\approx 1.14$, as compared to 
$R_c\approx 2.2~$fm and $F_b/T\approx 2.1$. At $T=123~$MeV, which 
corresponds to the critical temperature for the homogeneous case, 
the critical radius and $F_b/T$ diverge in the homogeneous computation, 
whereas $R_c\approx 35~$fm and $F_b/T\approx 394$ including 
inhomogeneities. The numbers above clearly indicate that the formation 
of critical bubbles is much less suppressed in the scenario with 
inhomogeneities, which will in principle accelerate the phase conversion 
process after the chiral transition.



In summary, the first inhomogeneity correction brings significant 
modifications to the form of the effective potential. Besides 
a general smoothening of the potential, the critical temperature 
moves upward and the hight of the barrier separating the symmetric 
and the broken phase vacua diminishes appreciably. As a direct 
consequence, the radius of the critical bubble goes down, as well 
as its free energy, and the process of nucleation is facilitated. 
This effect will compete with dissipation corrections that tend 
to slow down the process of phase conversion \cite{Fraga:2004hp}. 
Although the numbers presented above should be regarded as simple 
estimates, since they rely on a number of approximations, the 
qualitative behavior is clear. A more detailed analysis can be 
found in Ref. \cite{BT-ESF}


\begin{acknowledgments}
We thank D.G. Barci, H. Boschi-Filho, C.A.A. de Carvalho, T. Kodama, 
and especially A. Dumitru for fruitful discussions. 
This work was partially supported by CAPES, CNPq, FAPERJ and FUJB/UFRJ. 
\end{acknowledgments}





\end{document}